\documentclass[showpacs,aps,twocolumn,prl,floatfix]{revtex4-1}
\usepackage{amssymb}

\usepackage{amsmath}
\usepackage{graphicx}
\usepackage{bm}

\newcommand{\bea}{\begin{eqnarray}}
\newcommand{\eea}{\end{eqnarray}}

\begin{document}

\title{Tunable Charge and Spin Seebeck Effects in Magnetic Molecular Junctions}
\author{Pablo S. Cornaglia}
\affiliation{Centro At{\'{o}}mico Bariloche and Instituto Balseiro,
Comisi\'on Nacional de Energ\'{\i}a At\'omica, 8400 Bariloche, and CONICET,
Argentina}
\author{Gonzalo Usaj}
\affiliation{Centro At{\'{o}}mico Bariloche and Instituto Balseiro,
Comisi\'on Nacional de Energ\'{\i}a At\'omica, 8400 Bariloche, and CONICET,
Argentina}
\author{C. A. Balseiro}
\affiliation{Centro At{\'{o}}mico Bariloche and Instituto Balseiro,
Comisi\'on Nacional de Energ\'{\i}a At\'omica, 8400 Bariloche, and CONICET,
Argentina}

\begin{abstract}
We study the charge and spin Seebeck effects in a spin-$1$ molecular junction
as a function of temperature, applied magnetic field, and
magnetic anisotropy ($D$) using Wilson's numerical renormalization group. A
hard-axis magnetic anisotropy produces a large enhancement of the charge
Seebeck coefficient $S_c$ ($\sim k_B/|e|$)  whose value only depends on the residual interaction between quasiparticles in the low temperature Fermi-liquid regime.
In the underscreened spin-1 Kondo regime, the high sensitivity of the system
to magnetic fields makes it possible to observe a sizable value for the spin
Seebeck coefficient even for magnetic fields much smaller than the Kondo
temperature. Similar effects can be obtained in C$_{60}$ junctions where the control parameter, instead of $D$, is the gap between a singlet and a triplet molecular state. 
\end{abstract}
\date{\today}
\pacs{72.15.Qm, 72.15.Jf,  73.63.-b, 75.76.+j}

\maketitle

The Seebeck effect refers to the generation of a charge current (or
a voltage drop) by a tempe\-rature gradient applied across a metal \cite
{ashcroft1976solid}. The spin-Seebeck effect \cite
{Uchida2008,Bauer:2012fq,Lin:2012df}, concerns the thermal generation of
pure spin currents. Applications of the Seebeck and spin-Seebeck effects at
the nano-scale, with potential impact on a variety of new technologies,
could profit from the scalability and tunability properties of nano-devices
but still require a better understanding of thermopower effects in
nanostructures and of the effect of strong electron-electron correlations on
them. The recent experimental observation of the Seebeck effect in different
nano-structures, in particular in molecular junctions \cite{Reddy16032007}
and quantum dots (QDs) \cite{Scheibner:2005kh}, opened new routes to study
these phenomena. Here we show that junctions with spin-$1$ 
molecules, like the Co(tpy-SH)$_2$\ complex \cite{Parks2010}
or the C$_{60}$ buckyballs \cite{Roch:2008is}, give rise to a
large and controllable enhancement of the Seebeck and spin Seebeck
effects at low temperatures and low magnetic fields. We also show that
thermoelectric experiments in these systems give access to valuable
information on the residual interaction between quasiparticles in the low
temperature Fermi-liquid regime \cite{Nozieres:1974go}, information that
cannot be obtained by conventional techniques. This offers
the opportunity to experimentally study these effects on the smallest
possible lengths scales looking for the basic mechanisms of thermomagnetic
effects in strongly correlated systems \cite
{Oudovenko:2002ba,Costi2010,Rejec2012,Pfau:2012hc}---in particular in the
regime where energy transfer is governed by spin fluctuations.

\begin{figure}[t]
\centering
\includegraphics[width=0.45\textwidth,clip,angle=-90]{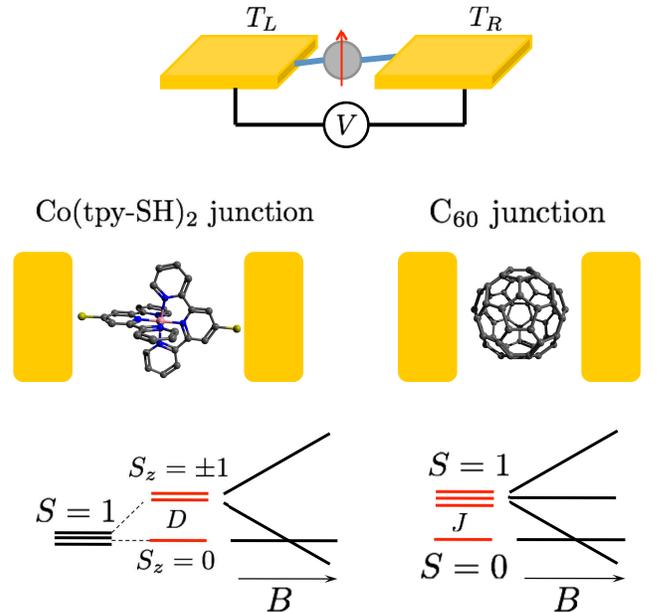}
\caption{Schematic representations of the Co(tpy-SH)$_{2}$ and C$_{60}$
molecular junctions. In the former, a distortion of the crystal field and/or
an external magnetic field induces the splitting of the $S=1$ state. In the C%
$_{60}$ case there is a singlet-triplet gap given by the exchange constant $J
$. }
\label{scheme}
\end{figure}

We start by describing the case of junctions with the organometallic
molecule Co(tpy-SH)$_{2}$ [see Fig. 1]. In this complex, the Co ion is
located in an approximately octahedral environment and is found to be in the
Co$^{1+}$ $3d^{8}$ configuration with a total spin $S=1$. When the molecule
is mechanically stretched, the local environment of the Co ion is distorted
and a magnetic anisotropy, with an easy plane, emerges due to the spin-orbit
interaction. 
As the molecular orbitals reproduce the magnetic structure of
the metallic ion, the molecular junction is modeled by the Hamiltonian 
$\hat{\mathcal{H}}=\hat{\mathcal{H}}_{M}+\hat{\mathcal{H}}_{V}+\hat{\mathcal{H}}_{l}$ where 
\begin{eqnarray}
\nonumber
\hat{\mathcal{H}}_{M}\!&=&\!\sum_{\ell =a,b}\left[ U\hat{n}_{\ell
\uparrow }\hat{n}_{\ell \downarrow }\!+\!\varepsilon(\hat{n}_{\ell
\uparrow }+\hat{n}_{\ell \downarrow })\right] +J\hat{\bm{S}}_{a}\cdot \hat{%
\bm{S}}_{b}\\
&&-g\mu _{B}BS_z+D\hat{S}_{z}^{2},
\end{eqnarray}
describes two degenerate orbitals ($a$ and $b$) of the molecule with
level energy $\varepsilon$ and Coulomb repulsion $U$, coupled through a spin exchange interaction $J$.  Here $\hat{\bm{S}}=\hat{\bm{S}}_{a}+\hat{\bm{S}}_{b}$ is the spin operator, $\hat{\bm{S}}_{\ell }=\sum_{\sigma \sigma ^{\prime }}d_{\ell \sigma}^{\dagger }\hat{\bm{\sigma}}_{\sigma \sigma ^{\prime }}d_{\ell \sigma }^{{}}$  where $\hat{\bm{\sigma}}_{\sigma \sigma ^{\prime }}$ the Pauli vector, and $\hat{n}_{\ell\sigma }\!=\!d_{\ell \sigma }^{\dagger }d_{\ell \sigma }^{{}}$ is the number operator. The molecular spin is coupled to an external magnetic field $B$, and $D$ is the strength of the magnetic anisotropy. In this case, the orbitals describe the Cobalt $e_g$ levels and $J\!<\!0$ is a Hund rule ferromagnetic coupling---in the $C_{60}$ case, $J\!>\!0$ is antiferromagnetic and $D=0$.
The hybridisation between the molecular orbitals and the leads' states is described by $\hat{%
\mathcal{H}}_{V}=\sum_{\bm{k}\sigma \alpha }\,V_{\bm{k}\alpha}\left(
d_{a\sigma }^{\dagger }c_{\bm{k}\sigma \alpha }^{{}}+c_{\bm{k}\sigma \alpha
}^{\dagger }d_{a\sigma }^{{}}\right) $, where $\alpha =L,R$ stands for the
left and right electrodes, respectively. Here we have assumed that a single
conduction channel is active in order to describe the underscreened Kondo
effect observed in Ref. \cite{Parks2010}, and we have considered, for simplicity, that only one of the molecular orbitals is coupled to the electrodes. 
The Hamiltonian of the metallic leads is given by $%
\hat{\mathcal{H}}_{l}=\sum_{\alpha \bm{k}\sigma }\varepsilon _{\bm{k}{\alpha
}}^{{}}\,c_{\bm{k}\sigma \alpha }^{\dagger }c_{\bm{k}\sigma \alpha }^{{}}$. 
We focus on the regime $\varepsilon<0$, and $U\gg |\varepsilon|$ where only the single and double occupied states of the molecule are relevant for the low--temperature physics.

The presence of a voltage ($V$) and temperature drop ($\Delta T$) across the
junction generates a charge current that in the linear response regime is
given by 
\begin{equation}
I_{c}=G\,V+GS_{c}\,\Delta T
\end{equation}
where the conductance $G$ and charge Seebeck coefficient $S_{c}$ are given
by 
\begin{equation}
G=\frac{e^{2}}{h}\,\mathcal{I}_{0}\,,\qquad S_{c}=-\frac{k_{B}}{|e|}\frac{%
\mathcal{I}_{1}}{k_{B}T\,\mathcal{I}_{0}}  \label{GandS}
\end{equation}
with $\mathcal{I}_{n}=\sum_{\sigma }\,\mathcal{I}_{n}^{\sigma }$ and $%
\mathcal{I}_{n}^{\sigma }=\int_{-\infty }^{\infty }\omega ^{n}\left( -\frac{%
\partial f(\omega )}{\partial \omega }\right) \mathcal{T}_{\sigma }(\omega
)\,\mathrm{d}\omega \,$. Here the quantity $\mathcal{T}_{\sigma }(\omega )$
describes the tunneling of spin-$\sigma $ electrons across the junction and
is given by\ $\mathcal{T}_{\sigma }(\omega )=\frac{4\Gamma _{L}\Gamma _{R}}{%
\Gamma _{L}+\Gamma _{R}}\rho _{\sigma }(\omega )$ where $\rho _{\sigma
}(\omega )$ is the spin dependent spectral density of the molecular state.
In the above expression $\Gamma _{\alpha }=\pi \rho _{\alpha }V_{\alpha }^{2}
$ is the contribution to the width of the molecular energy levels introduced
by the coupling with the $\alpha $ lead, $\rho _{\alpha }$
is the electronic density of states per spin of the electrodes at the Fermi
level and $f(\omega )$ is the Fermi function. In what follows we assume  $\rho_\alpha$ to be constant, set the Fermi energy to zero ($\varepsilon_F=0$) and choose the bandwidth of the leads $W$ as the unit of energy.

\begin{figure}[t]
\centering
\includegraphics[width=0.45\textwidth,clip]{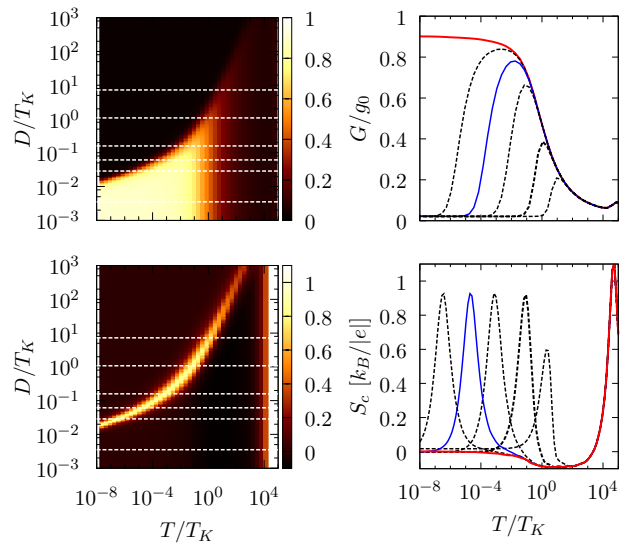}
\caption{Left panels: color maps of the conductance $G$ (top) and charge
Seebeck coefficient $S_{c}$ (bottom) as a function of the temperature and
the anisotropy constant $D$. For finite $D$, the conductance goes down to
very small values for $T<T_{K}^{\star }$ while the $S_{c}$ reaches its
maximum, $(k_{B}/|e|)\,\protect\pi /\protect\sqrt{3(1+\protect\gamma )}$,
for $T\lesssim T_{K}^{\star }$. Right panels: temperature dependence of $G$
and $S_{c}$ for different values of $D$, indicated in the color maps with
dashed lines. The conductance is given in units of  $g_{0}=(2e^{2}/h)\,4%
\Gamma _{L}\Gamma _{R}/(\Gamma _{L}+\Gamma _{R})^{2}$. 
}
\label{Hzero}
\end{figure}
The molecular spectral density is evaluated using the numerical
renormalization group (NRG), a non-perturbative technique known to give
excellent results for the transport integrals \cite{wilson_1975,Bulla2008}.
The total conductance and the charge Seebeck coefficient $S_{c}$ are shown
in Fig. \ref{Hzero} for different values of the anisotropy parameter $D$ and 
$B=0$. For the isotropic case ($D=0$) the molecular spin is partially
screened by the conduction electrons below a Kondo temperature $T_{K}$ (the
under-screened Kondo effect) with $T_{K}$ ranging between $1$K and $200$K
depending on the device \cite{Parks2010}. The ground state of the system is
a singular Fermi-liquid with a free (unscreened) spin $\frac{1}{2}$ \cite
{logan,Hewson2005Sing}. The development of Kondo correlations is associated
with a monotonic increase in the conductance as $T$ is lowered (top panels
Fig. \ref{Hzero}).\ With a spin anisotropy $0<D\leq k_{B}T_{K}$ , as the
temperature is lowered the conductance first increases following a universal
behavior, goes through a maximum and decreases at a cha\-rac\-te\-ris\-tic
temperature $T_{K}^{\star }=T_{K}\exp (-2\sqrt{k_{B}T_{K}/D})$ \cite
{Cornaglia2011}. This new temperature scale can be identified with a second
stage Kondo effect, induced by the magnetic anisotropy, in which the
remaining spin $\frac{1}{2}$ is screened. In this regime the ground state of
the system is a Fermi liquid.

\begin{figure}[t]
\centering
\includegraphics[width=0.4\textwidth,clip]{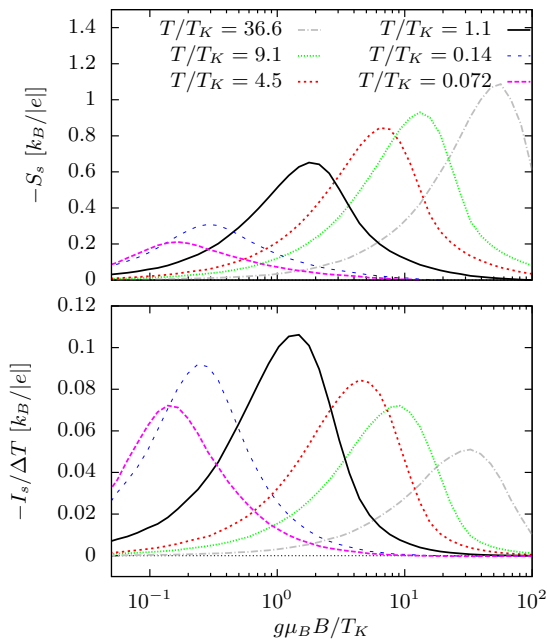}
\caption{Spin Seebeck coefficient $S_{s}$ and spin current $I_{s}$ for an
isotropic molecule ($D=0$). Top panel: $S_{s}=S_{\uparrow }-S_{\downarrow }$
as a function of the magnetic field for different temperatures. Note that $%
S_{s}$ is significantly large even for $g\mu _{B}B\ll k_{B}T_{K}$.
The maximum of $S_{s}$ occurs for $g\mu _{B}B\sim k_{B}T$. Bottom
panel: Spin current as a function of $g\mu _{B}B$ in the zero charge
current condition ($I_{c}=0$).}
\label{D0}
\end{figure}

\begin{figure*}[t]
\centering
\includegraphics[width=0.45\textwidth,clip]{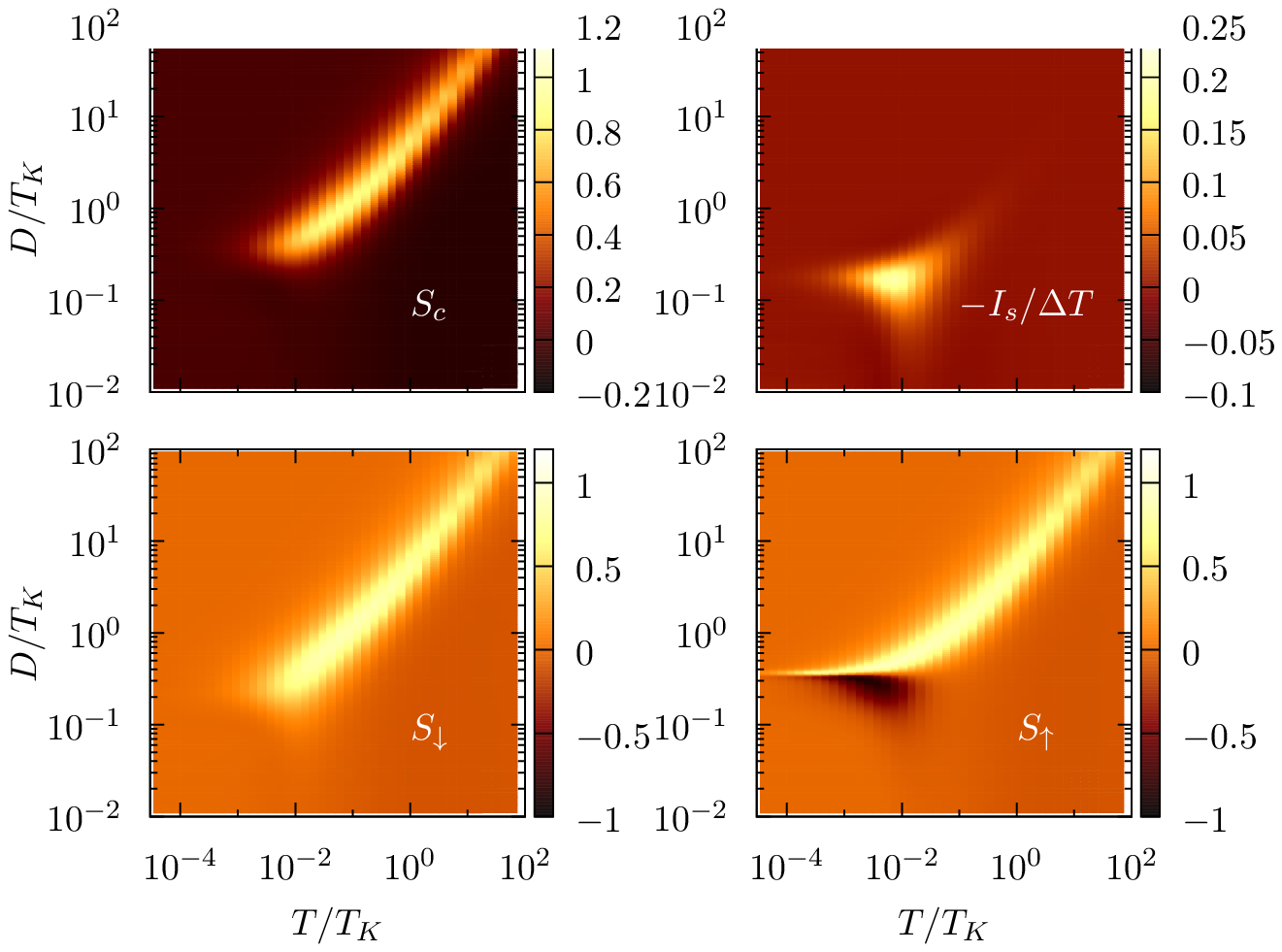} %
\includegraphics[width=0.48\textwidth,clip]{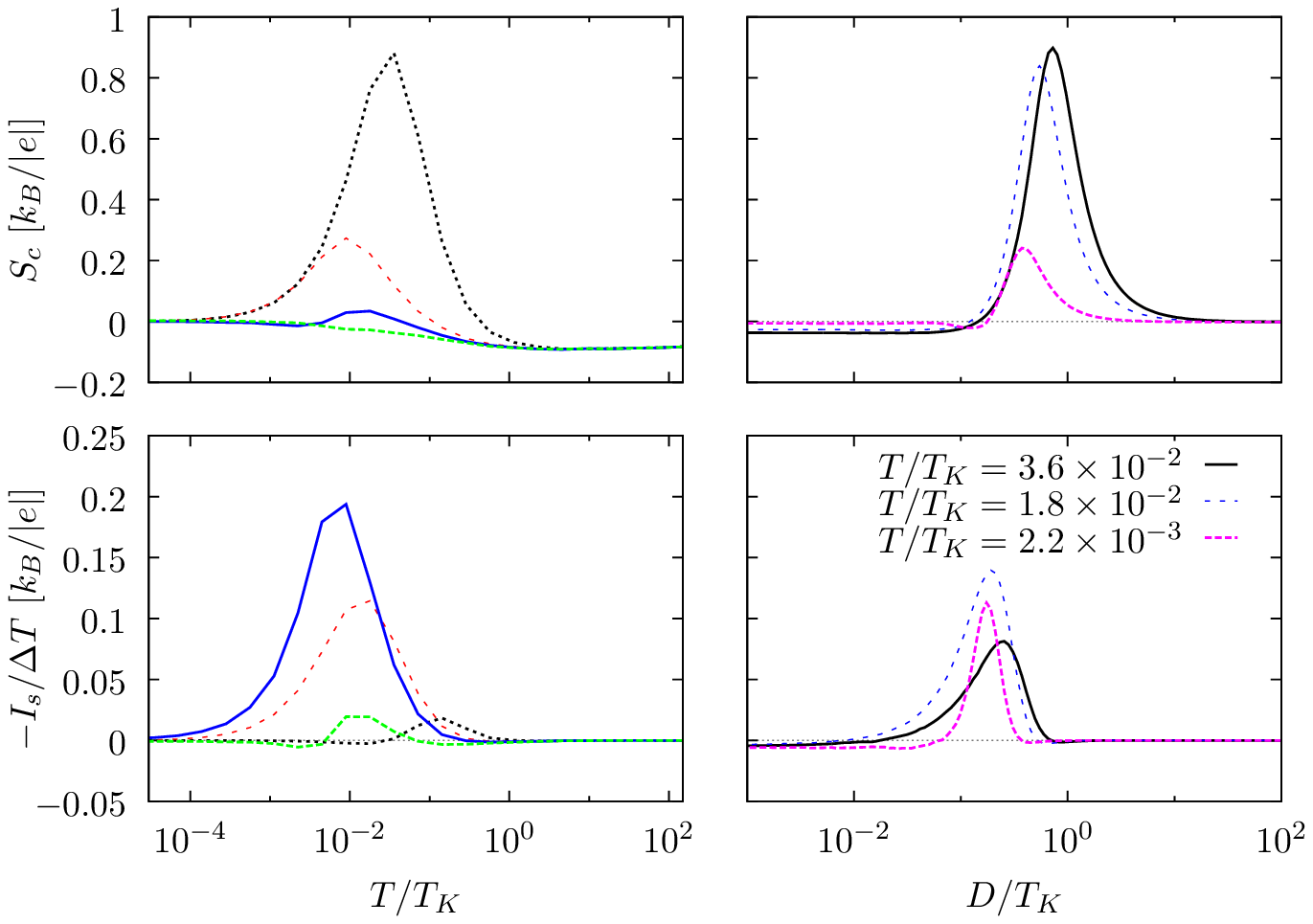}
\caption{Left column panels: color maps of the Seebeck coefficient ($S_{c}$%
), the spin current ($I_{s}$) and the spin dependent Seebeck coefficients ($%
S_{\sigma }$) as a function of the temperature and the anisotropy
constant $D$. In all cases $g\mu _{B}B=0.01k_{B}T_{K}$ Note that $%
S_{c}$ is strongly reduced for $D/T_{K}\lesssim 0.5$ that corresponds to $g%
\mu _{B}B\lesssim k_{B}T_{K}^\star$ and that $I_{s}$ reaches its
maximum value in the parameters region where $S_{\downarrow }$ changes sign.
Notice that, as $D$ decreases, the $S_\downarrow$ peak moves down to very
low $T$ before changing its sign. Right column panels: Temperature and
anisotropy dependence of $S_{c}$ and $I_{s}$. Note that $S_{c}$ is
suppressed in the region where $I_{s}$ is different from zero.}
\label{Dvar}
\end{figure*}

Within this scenario, that has been confirmed by experiments and theory, the
Seebeck coefficient shows novel features (bottom panels Fig. \ref{Hzero}).
For $D=0$ it develops a large peak at very high temperatures (unphysical for
magnetic molecules) and changes sign at a lower temperature. This behavior
is similar to what is obtained for spin-$\frac{1}{2}$ quantum dots \cite
{Costi2010,Rejec2012}. For the stretched ($D>0$) molecule \footnote{%
A small distortion in the unstretched molecule would be observable only if $%
T_{K}^{\star }$ is large enough. Experimentally, anisotropy effects in the
conductance are observed only for moderate or large stretching, a fact that
is consistent with the exponential dependence of $T_{K}^{\star }$ on the
parameter $D$. We have assume, without loss of generality, that $D=0$
corresponds to the unstretched molecule.}, a second large peak develops at $%
T\lesssim T_{K}^{\star }$. This low-$T$ peak is controlled by spin
fluctuations and it is strongly dependent on the spin anisotropy. For small
anisotropy ($D<k_{B}T_{K}$), the two characteristic energy scales $%
T_{K}^{\star }$ and $T_{K}$ are far apart and $S_{c}$ is a function of $%
T/T_{K}^{\star }$ at low temperature. For larger anisotropies, the scaling
breaks down and the maximum value of $S_{c}$ decreases. In the scaling
region, and for $T/T_{K}^{\star }<1$, the system is a Fermi-liquid and $S_{c}
$ can be described by a simple analytical expression. In Fermi-liquids, the
residual interaction between quasiparticles leads to a self-energy with an
imaginary part proportional to $(\omega ^{2}+\pi ^{2}T^{2})$ 
\cite{Nozieres:1974go,pines1989theory}. U\-sing this
result to estimate the spectral density $\rho _{\sigma }(\omega )$ we obtain 
\cite{supp}, 
\begin{equation}
S_{c}=\frac{k_{B}}{\left| e\right| }\frac{2\pi ^{2}(k_{B}T/\omega _{0})}{\pi
^{2}(1+\gamma )(k_{B}T/\omega _{0})^{2}+3}\,.  \label{Sc}
\end{equation}
Here $\omega _{0}=k_{B}T_{K}^{\star }\sin (\pi n/2)$, $n$ is the total
occupation of the molecular orbitals and $\gamma >0$ measures the strength
of the quasiparticle interaction. Notably, the maximum of the Seebeck
coefficient $S_{c}^{\mathrm{max}}=(k_{B}/\left| e\right| )\pi /\sqrt{%
3(1+\gamma )}$ depends only on universal constants and the parameter $\gamma 
$ that reduces the maximum value of the thermopower---the coupling to the
leads, $D$ and other parameters that are both experimentally hard to
determine and sample dependent, do not affect the value of $S_{c}^{\mathrm{%
max}}$. The temperature behavior of the Seebeck coefficient is one of the
central results of this work. Its measurement provides a direct access to
the parameter $\gamma $ and can be used to quantitatively test Nozi\`{e}res
local Fermi liquid ideas \cite{Nozieres:1974go}. 

The fitting of the numerical results with Eq. (\ref{Sc}) is excellent (see 
\cite{supp}) and gives $\gamma \simeq 2.8$ for the parameters used in our
simulation. Therefore, in a junction with $T_{K}=100$K, the Seebeck
coefficient could be tuned to have a peak of the order of $k_{B}/|e|\sim
100\mu V/K$ at a temperature in the range of $0$-$50$K.

Let us now focus on the effect of an external magnetic field. There are now
two control parameters, the magnetic field $B$ and the anisotropy $D$,
leading to different magneto-thermal regimes for the junction. Defining the
contribution to the charge current of the spin-$\sigma $ electrons as $%
I_{\sigma }=G_{\sigma }V+G_{\sigma }S_{\sigma }\Delta T$, where $G_{\sigma }$
is the $\sigma $-contribution to the total conductance [cf. Eq. (\ref{GandS}%
)], the spin dependent Seebeck coefficient results, 
\begin{equation}
S_{\sigma }=-\frac{k_{B}}{|e|}\frac{\mathcal{I}_{1}^{\sigma }}{k_{B}T\,%
\mathcal{I}_{0}^{\sigma }}\,.
\end{equation}
If the condition $I_{c}=0$ is set, a voltage drop across the junction will
develop as before but in addition, the temperature difference $\Delta T$
will also generate a pure spin current $I_{s}=(I_{\uparrow }-I_{\downarrow
})/|e|$ given by 
\begin{equation}
I_{s}=(S_{\uparrow }-S_{\downarrow })\frac{\Delta T}{|e|\mathcal{R}}
\end{equation}
with $\mathcal{R}=\sum_{\sigma }G_{\sigma }^{-1}$. In this units $I_{s}$
gives the number of spins per second flowing though the junction. Note that
with these definitions $S_{c}=\sum_{\sigma }(G_{\sigma }/G)S_{\sigma }$. For 
$D=0$, the under-screened Kondo regime, the system behaves as a singular
Fermi-liquid with a divergent spin susceptibility as $T\rightarrow 0$ \cite
{Cornaglia2011} and $\rho (\omega )$ shows a Kondo peak with a singular
behavior at the Fermi energy $E_{\text{F}}$. As the ground state of the
system contains a free $\frac{1}{2}$-spin, the presence of a magnetic field
is a relevant perturbation in the renormalization group sense. Consequently,
even a very small magnetic field is able to spin polarize the molecule and
split the Kondo peak in $\rho (\omega )$ (see \cite{supp}). The magnetic
field dependence of $S_{s}=S_{\uparrow }-S_{\downarrow }$ and $I_{s}$ is
shown on Fig. \ref{D0}. This spin polarization of the molecule results in
the appearance of a pure spin current even for $g\mu _{B}B\ll k_{B}T_{K}$
(but $g\mu _{B}B\sim k_{B}T$), in clear contrast with the $S=\frac{1}{2}$
case (Anderson model) where $g\mu _{B}B\sim k_{B}T_{K}$ is required in order
to observe a sizable spin polarizing effect \cite{Rejec2012}.

Similar results are obtained for the stretched molecule only if $g\mu
_{B}B\gtrsim k_{B}T_{K}^{\star }$. For $g\mu _{B}B<k_{B}T_{K}^{\star }$ the
coefficient $S_{s}$ is small and a large charge Seebeck peak is recovered
(see Fig 3) This behavior shows that with some spin anisotropy ($D\neq 0$)
the voltage drop $V$ at the junction is very sensitive to the external field
and can evolve from a large value (of the order of $k_{B}\Delta T/|e|$) for $%
B=0$ to zero for $g\mu _{B}B\gtrsim k_{B}T_{K}^{\star }$ \cite{Walter:2011br}%
. 
The low temperature behavior of the spin dependent Seebeck coefficients $%
S_{\sigma }$ can also be described using a local Fermi liquid theory. We
obtain expressions similar to Eq. (\ref{Sc}) where now $\omega _{0}$ is
replaced by $\omega _{0\sigma }=k_{B}T_{K}^{\star }\sin (\pi n_{\sigma })$,
with $n_{\sigma }\equiv n/2-\sigma m/2$ the total number of spin-$\sigma $
electrons in the molecular orbitals. As $n\simeq 2$, a small magnetization $%
m $ can change the sign of one of the energies $\omega _{0\sigma }$ and
consequently the sign of one of the spin dependent Seebeck coefficients $%
S_{\sigma }$ leading to a large spin current.

C$_{60}$ junctions present similar thermopower properties. A C$_{60}$
molecule embedded in a junction may have a singlet $S=0$ state and a triplet 
$S=1$ state separated by a small singlet-triplet gap $J$ that can be
controlled by gate voltages \cite{Roch:2008is}. When the singlet state lies
just below the triplet state the system shows a two stage Kondo effect with
a Kondo temperature $T_{K}$, of the order of $4$K in \cite{Roch:2009ui} ,
and a second stage temperature given by $T_{K}^{\star }\propto T_{K}\exp
(-k_{B}T_{K}/J)$ \cite{cornaglia2005,Roch:2009ui}. While the global behavior
of the conductance is similar to that of the anisotropic spin-$1$ case, its
specific high $T$-dependence is different. By contrast, the temperature
dependence of $S_c$ for $T$ $\lesssim k_{B}T_{K}^{\star }$ is the same as
the one obtained for the Co-complex and it is well described by Eq (\ref{Sc}%
) with a par\-ti\-cu\-lar value for $\gamma $ that depends on the residual
interaction in the $C_{60}$ case. If the  triplet is the ground state, and for $J>T_{K}$,
the system is described by the isotropic under-screened $S=1$ Kondo model
and the physical behavior corresponds to the $D=0$ case discussed above. It
is worth mentioning that other systems, like an artificial molecule build
with a T-shape double Quantum Dot (2QD) may also show some similar effects.
In these two cases, however, the Kondo temperatures are usually very small
making it more difficult to experimentally observe them. In this
sense, the stretched Co complex presents advantages over these systems that
make it an excellent candidate to observe high charge and spin Seebeck
effects at low or moderate fields and temperatures.

In summary we have presented a study of the charge and spin Seebeck effects
for real molecular junctions that combines the $S=1$ underscreened Kondo
effect together with a controllable spin anisotropy $D$ or a single-triplet
gap $J$. We found that these systems presents a large charge Seebeck
coefficient $S_{c}$, of the order of $k_{B}/|e|$, at a temperature that can
be controlled either by stretching or gating the molecules. Quite
remarkably, the maximum value of $S_{c}$ in the low $T$ regime, $%
(k_{B}/|e|)\pi /\sqrt{3(1+\gamma )}$, depends only on the Fermi liquid
properties of the system (characterized by $\gamma $). It is important to
emphasize that this low $T$ behavior is fully controlled by spin
fluctuations. In the presence of a magnetic field these systems are able to
filter the spin flow in a regime where Kondo correlations are fully
developed and present moderate values of $S_{s}$ and $I_{s}$ even for
magnetic fields such that the Zeeman energy is much smaller than the Kondo
energy scale.

We acknowledge financial support from PICTs 06-483, 2008-2236 and
Bicentenario 2010-1060 from ANPCyT and PIP 11220080101821 from CONICET,
Argentina. 


%

\end{document}